\documentclass[aps,prl,twocolumn,amssymb,showpacs,superscriptaddress,10pt, longbibliography]{revtex4-1}

\usepackage{acronym}
\usepackage{amsfonts}
\usepackage{amsmath}
\usepackage{amssymb}
\usepackage{bm}
\usepackage{color}
\usepackage{latexsym}
\usepackage{dcolumn}
\usepackage{epsfig}
\usepackage{graphicx}
\usepackage{hyperref}
\usepackage{indentfirst}
\usepackage[latin1]{inputenc}
\usepackage{latexsym}
\usepackage{mathrsfs}
\usepackage{multirow}
\usepackage{rotating}
\usepackage{subfigure}
\usepackage{tabls}
\usepackage{upgreek}
\usepackage{verbatim}

\hypersetup{
    pdfauthor={Santos-Oliv\'an, Daniel and Sopuerta, Carlos F.},
    pdftitle={New Features of Gravitational Collapse in Anti-de Sitter Spacetimes}}

\newcommand{\be}{\begin{equation}}
\newcommand{\ee}{\end{equation}}
\newcommand{\bea}{\begin{eqnarray}}
\newcommand{\eea}{\end{eqnarray}}

\newcommand{\onehalf}{\textstyle{\frac{1}{2}}}
\newcommand{\met}{\mbox{g}}

\begin{document}
\title{New Features of Gravitational Collapse in Anti-de Sitter Spacetimes}

\author{Daniel Santos-Oliv\'an}
\affiliation{Institut de Ci\`encies de l'Espai (CSIC-IEEC), 
Campus UAB, Carrer de Can Magrans s/n, 08193 Cerdanyola del Vall\`es, Spain.}

\author{Carlos F.~Sopuerta}
\affiliation{Institut de Ci\`encies de l'Espai (CSIC-IEEC), 
Campus UAB, Carrer de Can Magrans s/n, 08193 Cerdanyola del Vall\`es, Spain.}

\date{\today}


\begin{abstract} 
Gravitational collapse of a massless scalar field in spherically-symmetric anti-de Sitter (AdS) spacetimes presents a new phenomenology with a series of critical points whose dynamics is discretely self-similar as in the asymptotically flat case. 
Each critical point is the limit of a branch of scalar field configurations that have bounced off the AdS boundary a fixed number of times before forming an apparent horizon.  
We present results from a numerical study that focus on the interfaces between branches. 
We find that there is a mass gap between branches and that subcritical configurations near the critical point form black holes with an apparent horizon mass that follows a power law of the form $M_{\mathrm{AH}}-M_{g} \propto (p_{c}-p)^{\xi}$, where $M_g$ is the mass gap and the exponent $\xi\simeq 0.7$ appears to be universal.
\end{abstract}
 

\maketitle

\emph{Introduction.}---The discovery of critical phenomena in the general relativistic gravitational collapse by Choptuik~\cite{Choptuik:1992jv} remains nowadays as one of the main triumphs of numerical relativity (see, Refs.~\cite{Gundlach:2002sx,Gundlach:2007gc} for reviews).  Following a suggestion by Christodoulou~\cite{Christodoulou:1986zr} on whether gravitational collapse can form a black hole (BH) with arbitrarily small mass, Choptuik investigated the gravitational collapse of a massless scalar field in spherically symmetric asymptoticall -flat spacetimes at the threshold between BH formation and dispersion, with flat spacetime as the end state. It was found that such system exhibits critical phenomena with the critical solution being discretely self-similar (type II critical behavior) and such that supercritical configurations form BHs with masses that follow a power-law near the vicinity of the critical solution: $M_{\rm BH} \propto (p-p_{c})^{\gamma}$. Here $p$ parameterizes a one-parameter family of initial data, with the critical solution located at $p=p_{c}$ and $\gamma\simeq 0.37$, the scaling exponent, is universal, i.e. the same for any family of initial configurations.

The situation in asymptotically anti-de Sitter (AAdS) spacetimes is very different.  With the growing interest in the holographic methods provided by the anti-de Sitter/conformal field theory (AdS/CFT) correspondence~\cite{Maldacena:1997re} (see also, e.g. Ref.~\cite{Witten:1998qj}), the dynamics of critical collapse and the stability of AdS have been studied in detail in the last years.  This has produced a number of developments in numerical relativity to deal with AAdS spacetimes and the AdS/CFT correspondence (see, e.g. Refs.~\cite{Cardoso:2012qm,Cardoso:2014uka}). The first numerical studies of gravitational collapse in AAdS spacetimes were done in $2+1$ spacetime dimensions~\cite{Husain:2000vm,Pretorius:2000yu} (see also, Ref.~\cite{Jalmuzna:2015hoa}), where the dynamics is qualitatively different from the case of $d+1$ (with $d\geq 3$) dimensions.  In $d\geq 3$ it has been shown that massless spherically symmetric scalar fields exhibit an instability~\cite{Bizon:2011gg,Jalmuzna:2011qw}, called the AdS {\em turbulent} instability, consisting in the formation of an apparent horizon (AH) for general families of initial configurations and independent of the initial energy.  This can be understood in terms of the AdS causal structure since lightlike signals can reach the AdS timelike boundary in a finite proper time.  Then, an initial configuration with arbitrary energy bounces repeatedly off the AdS boundary while the nonlinearity of gravity transfers energy from long wavelength to short wavelength modes (and hence the name {\em turbulent}), until the profile becomes sharp enough to form an AH.  The boxlike structure of AdS spacetime is fundamental as shown by studies in flat spacetime with an artificial boundary at a finite distance where the same type of instability appears~\cite{Maliborski:2012gx}. This shows the rich phenomenology of AdS but there are still open questions related to the existence of stable configurations~\cite{Maliborski:2013jca, Buchel:2013uba} and the fate of gravitational collapse.

\begin{figure}[b]
\begin{center}
\includegraphics[width=0.49\textwidth]{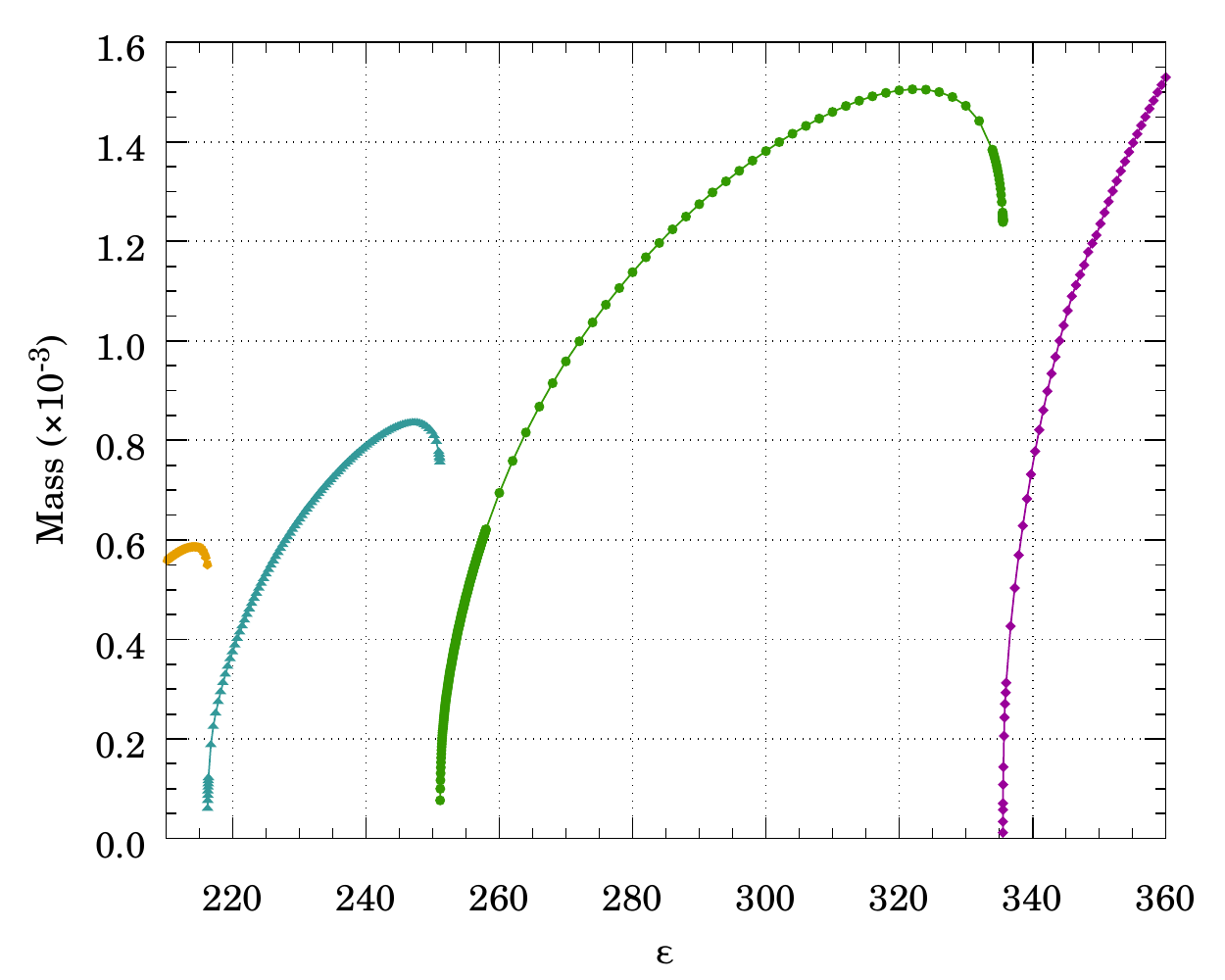}
\caption{AH Mass as a function of the amplitude $\varepsilon$ of the scalar field pulses of Eq.~(\ref{initial-configurations}) with $\sigma = 0.05$. We show the branches for (from right to left) zero, one, two, and three bounces. \label{plot_2d_general}}
\end{center}
\end{figure}

In this Letter, we investigate new features of gravitational collapse in AAdS spacetimes near the threshold of AH formation.  The current picture is well represented by Fig.~\ref{plot_2d_general}, where the AH mass with respect to the amplitude of the initial scalar field profile is shown, as in Ref.~\cite{Bizon:2011gg}.  Each branch corresponds to AH formation after a certain number of bounces off the AdS boundary, with the first branch from the right corresponding to direct collapse. By using a Cauchy-characteristic numerical code~\cite{Santos:2015ss}, that allows us to get quite close to the AH formation, we have studied the interface between these branches 
reaching the following conclusions: First, we confirm the conclusion of Ref.~\cite{Bizon:2011gg} that at the bottom of each branch there is type II critical behavior~\cite{Gundlach:2002sx,Gundlach:2007gc} with the same universal exponent for the scaling of the mass as in the asymptotically-flat case~\cite{Choptuik:1992jv} by fitting a large number of simulations near the critical point. Second, we corroborate that there is a mass gap between branches~\cite{Bizon:2011gg}, that is, those configurations that are close to form an AH with very small mass in a given branch but disperse and collapse after bouncing off the AdS boundary form black holes with a minimum mass, estimating the value for several branches.  Third, we find that in the upper part of a branch the configurations form BHs with masses following a power law of the type $M_{\mathrm{AH}}-M_{g} \propto (p_{c}-p)^{\xi}$, with $M_{g}$ being the minimum mass and $\xi\simeq 0.7$ is a universal exponent in the sense that it is the same for all branches and families of initial configurations.

\emph{Numerical construction of scalar-field AAdS spacetimes.}---The Einstein-Klein-Gordon (EKG) equations for a self-gravitating, real massless scalar field $\phi$ in a $3+1$ AAdS spacetime are~\footnote{We use units in which $c=1$ and $4\pi G = 1$. Numerical values of masses are given in units of $\ell$}:
\begin{equation} 
G_{\mu\nu} + \Lambda\, \met_{\mu\nu} = 2 \left( \phi_{;\mu}\phi_{;\nu} - \onehalf \met_{\mu\nu}\phi_{;\alpha}\phi^{;\alpha}  \right),  
\;
\phi_{;\mu}{}^{\mu} = 0\,, \label{ekgeqs} 
\end{equation}
where $\met_{\mu\nu}$ is the spacetime metric, $G^{}_{\mu\nu}$ is the Einstein tensor, $\Lambda$ is the (negative) cosmological constant ($\Lambda=-3/\ell^{2}$ and $\ell$ is the AdS length scale), and a semicolon denotes covariant differentiation. Most numerical approaches to solve numerically these equations in spherical symmetry use a Cauchy-evolution scheme, in contrast with the asymptotically-flat case, where a characteristic scheme like the one introduced in Refs.~\cite{Goldwirth:1987nu,Garfinkle:1994jb} is also commonly used.  The main reason is that the grid points in the characteristic formulation move towards the origin along ingoing null geodesics (the characteristics) and hence we cannot follow the scalar field bounces off the AdS boundary.  However, a characteristic scheme can bring us much closer to the AH formation than the Cauchy evolution. Here we use a hybrid scheme where the evolution of the EKG equations is initially followed using a Cauchy-evolution scheme, following the different bounces until the last one before AH formation. The metric for the Cauchy evolution is~\cite{Bizon:2011gg}:
\begin{equation}
ds^{2} = \frac{\ell^{2}}{\cos^{2}x}\left( - A {e}^{-2\delta}\,dt^{2} + \frac{dx^{2}}{A} + \sin^{2}x\, d\Omega^{2}_{2} \right)\,, \label{aadstx}
\end{equation}
where the metric functions $A$ and $\delta$ ($A=1$ and $\delta=0$ in pure AdS) depend on the time $t$ and the radial compactified coordinate $x$ that goes from the origin ($x=0$) to the AdS boundary ($x=\pi/2$) and $d\Omega^{2}_{2}$ is the unit $2$-sphere metric.  The Cauchy-type evolution of the EKG equations uses a multidomain pseudospectral method (with an adaptive algorithm to distribute the domains according to resolution needs) in combination with a Runge-Kutta 4 time-step algorithm.  After the last bounce, we use the Cauchy-evolution information to construct initial data on a null surface $u=$const whose points are evolved along ingoing null geodesics using a characteristic scheme. The metric for the characteristic evolution is:
\begin{equation}
ds^{2} = -g\bar{g}\,du^{2} - 2 g \,dudr + r^{2}\, d\Omega^{2}_{2}\,, 
\label{charmet}
\end{equation}
where the metric functions $g$ and $\bar g$ depend on $r=\ell\tan x$ and on the null coordinate $u$.  The transition from the Cauchy to the characteristic evolution is a key point of this method (details will be given elsewhere~\cite{Santos:2015ss}). An important ingredient of the transition is the following relation between the different metric functions: $A = \bar{g}g^{-1}( 1 + r^2/\ell^2)^{-1}$.
Then, we evolve the EKG equations with the characteristic scheme to approach as much as possible AH formation, which happens near the origin where $A\rightarrow 0$ ($\bar{g}/g\rightarrow 0$).  We typically reach values in the range $A\sim 10^{-12}-10^{-8}$.

\begin{figure}[b]
\begin{center}
\includegraphics[width=0.49\textwidth]{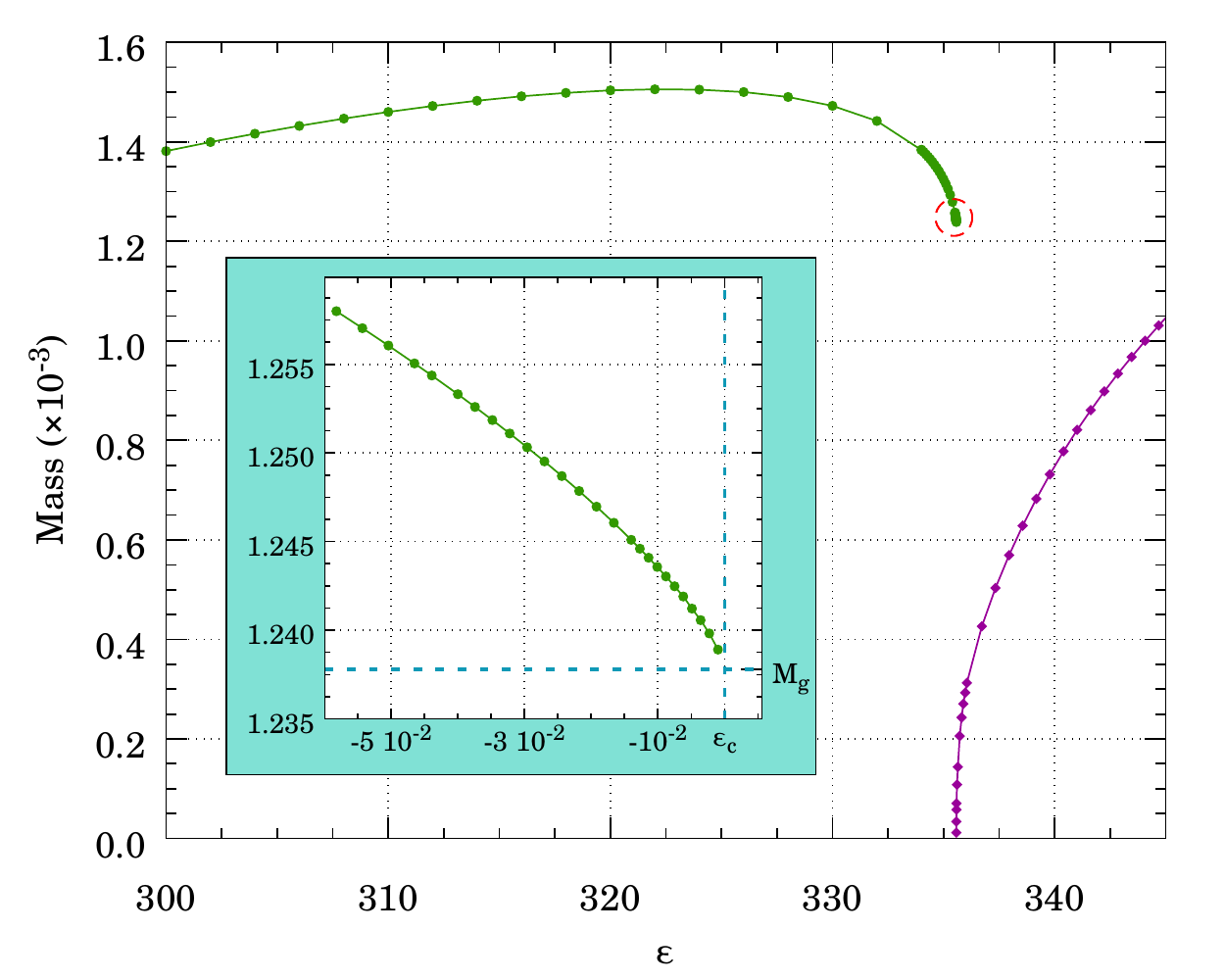}  
\caption{Interface between the zero and one bounce branches. The zoommed-in area shows the behavior as the critical point is approached from the one bounce branch. \label{plot_2d_massgap}}
\end{center}
\end{figure}

\begin{figure}[t]
\centering
\includegraphics[width=0.49\textwidth]{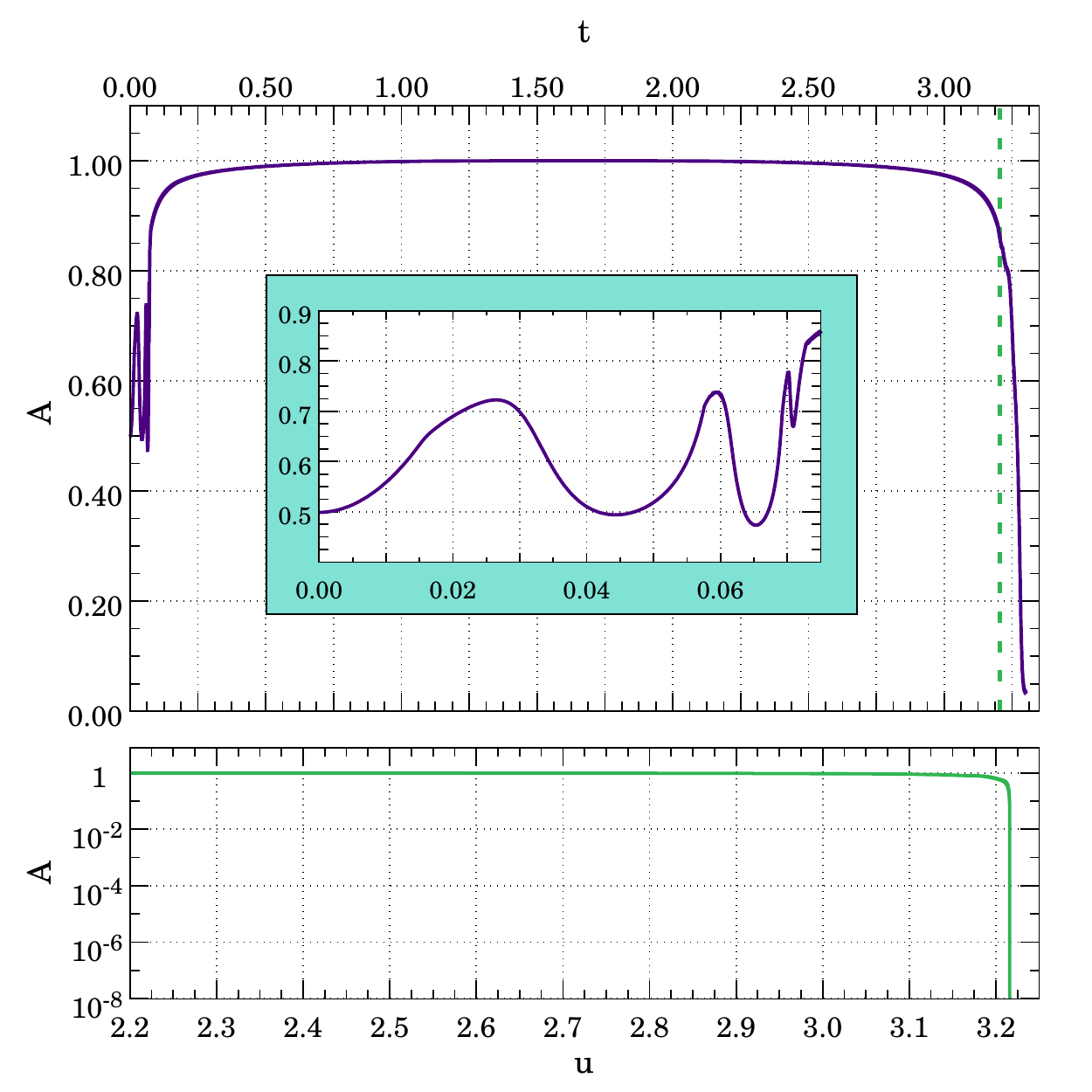}
\caption{Evolution of $A$ in the case of AH formation after one bounce. The upper panel shows the Cauchy evolution.  The vertical dashed line indicates the last time used for the transition to the characteristic evolution (lower panel). \label{plot_fit_evolA}}
\end{figure}

In order to explore the interface between branches (see Fig.~\ref{plot_2d_massgap}), near the critical point, very demanding numerics is required due to two main difficulties: (i) During the Cauchy evolution the nonlinearity of Einstein's equations produces the cascade of long-wavelength to short-wavelength modes making the scalar field profile sharper.  Near subcritical configurations that are very close to form an AH with small mass in a given branch but instead survive, travel to the AdS boundary, and bounce back towards the origin where they finally collapse.  Following the last trip to the AdS boundary and back is numerically very challenging since the scalar field has developed very sharp features that we need to propagate with high accuracy all the way to the AdS boundary and back.  (ii) In the characteristic evolution our grid points move towards the origin following ingoing null geodesics. In the case of AAdS spacetimes, and due to the negative cosmological constant that acts like a spring attached to the origin, even points near the AdS boundary come close to the origin by the time of AH formation and this puts stringent constraints on the construction of the initial null surface and its evolution with the characteristic scheme.

\emph{Numerical experiments and results}---We have performed a series of evolutions of the EKG system for the following two-parameter family of scalar field initial configurations (similar to those used in Ref.~\cite{Bizon:2011gg}):
\begin{equation}
\partial_x \phi = 0, \quad
\frac{\partial_t {\phi}}{A e^{-\delta}} = -\varepsilon  \,\cos x\, \exp\left(-\frac{4 \tan^2 x}{\pi^2 \sigma^2}\right), 
\label{initial-configurations}
\end{equation}
where the amplitude $\varepsilon$ and the width $\sigma$ are the parameters we change in our simulations.

For fixed $\sigma$, configurations with high enough amplitude collapse directly as in the asymptotically-flat case~\cite{Choptuik:1992jv}. We focus on near subcritical configurations for the one bounce branch. In Fig.~\ref{plot_fit_evolA} we show the evolution of the metric function $A$ for one such configuration. 
Initially, the evolution is close to collapse (and to the critical solution) and exhibits oscillations typical of type II critical phenomena.
After the trip to the AdS boundary and back, the scalar field collapses straightaway.  In Fig~\ref{plot_2d_general}, we show the AH mass in terms of the initial scalar field amplitude $\varepsilon$ for a constant width $\sigma=0.05$, in which the collapse happens either directly (purple) or after a few bounces.  In Fig.~\ref{plot_2d_massgap} we look at the interface between the branch of direct collapse and the branch of collapse after one bounce.  We find that BHs formed after one bounce have a minimum AH mass, $M_{g}$, and that the AH mass follows a power law of the type $M_{\mathrm{AH}}-M_{g} \propto (p_c - p)^\xi$.  An argument about why there are no points with arbitrarily small mass in this side of the branch is that the subcritical configurations have to travel to the AdS boundary and come back, suffering a {\em finite compression} due to the nonlinearity (a key ingredient for the {\em turbulent} instability), that makes them to form BHs with an initial minimum mass.

\begin{figure*}[t]
\centering
\includegraphics[width=0.3\textwidth]{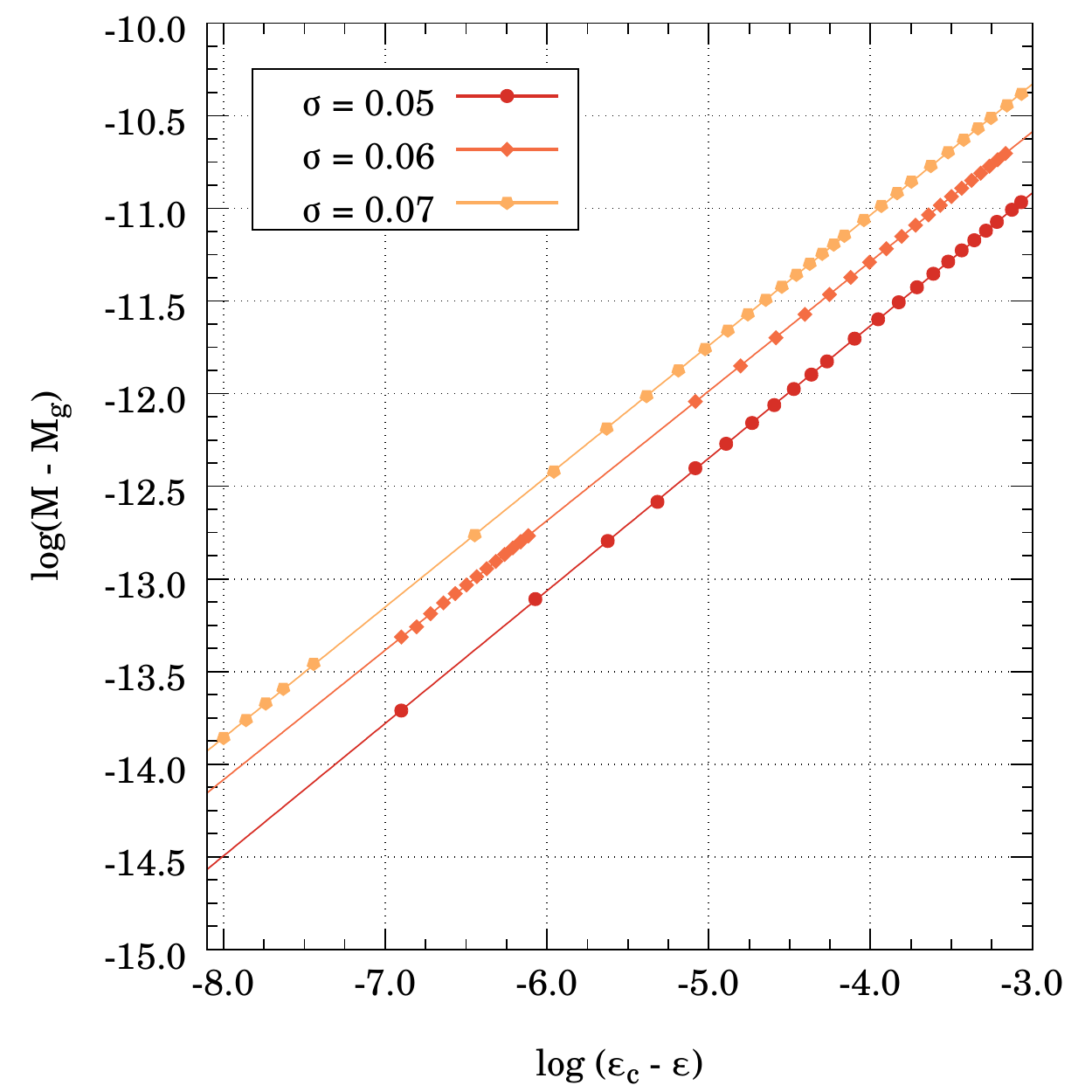}
\includegraphics[width=0.3\textwidth]{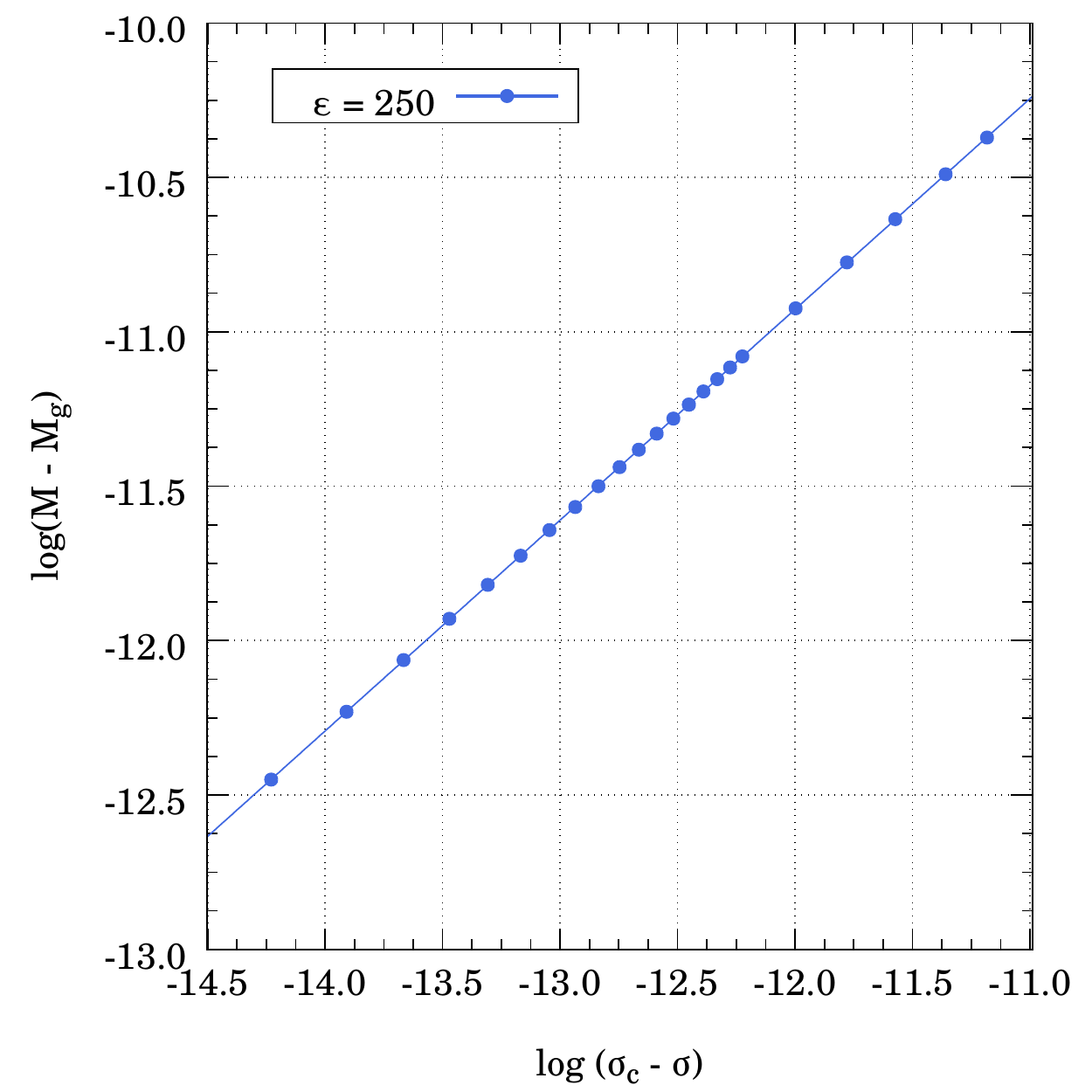}
\includegraphics[width=0.3\textwidth]{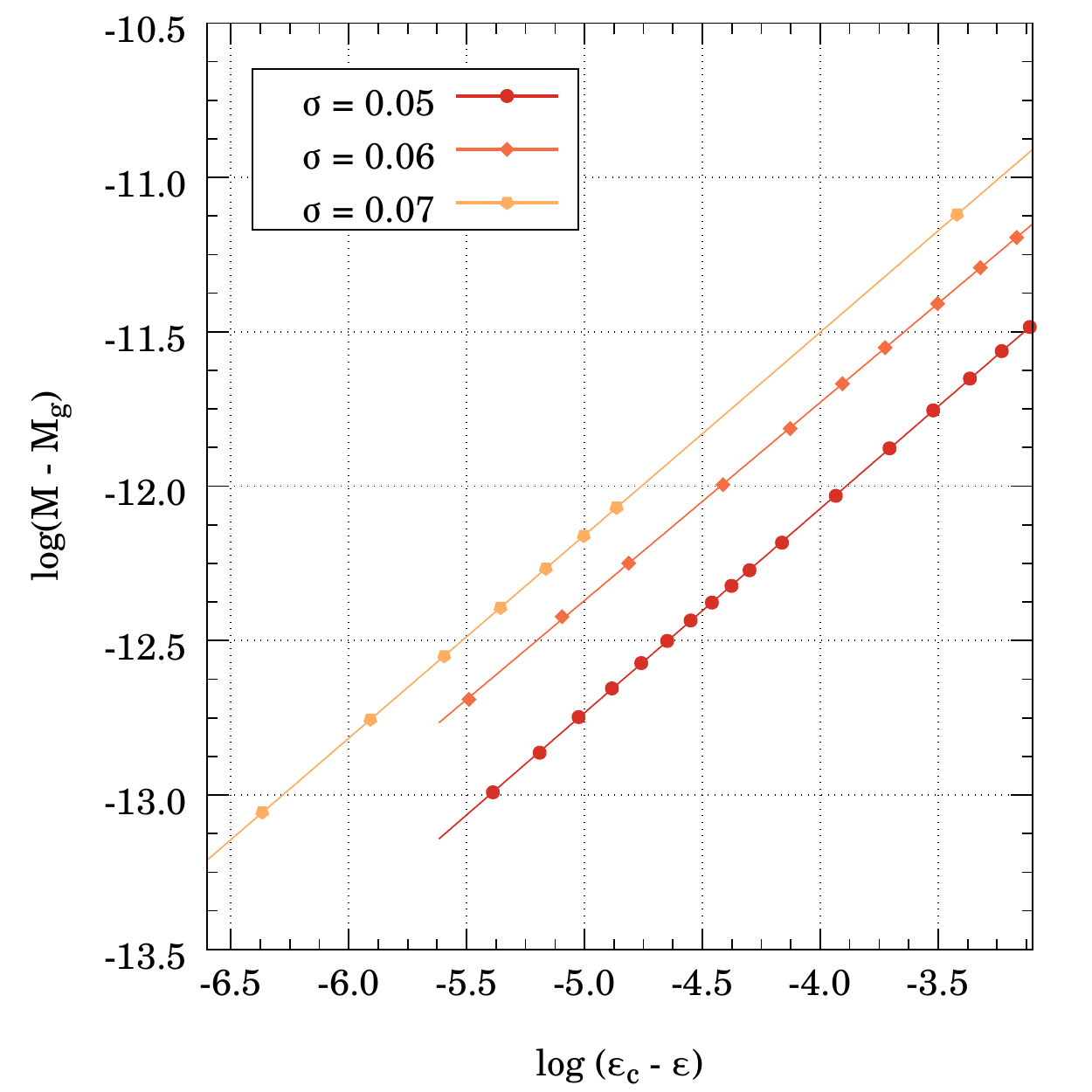}
\caption{Fitting plots for: (left) simulations for the one-bounce branch and constant scalar field width; (center) simulations for the one bounce branch and fixed amplitude; (right) simulations for the two-bounce branch and constant scalar field width.  \label{massgap-fits}}
\end{figure*}

\renewcommand{\arraystretch}{1.2}
\renewcommand{\tabcolsep}{8pt}
\begin{table*}[t]
\begin{center}
\begin{tabular}{c c| c c c }
\hline\hline
Gap number &  Fixed parameter & Critical value~($p_c$)   & Mass gap~($M_g$)     &  Exponent~($\xi$) \\
\hline
$1$ & $\sigma = 0.05$ 	&	$\varepsilon_c = 335.572 \pm 0.002$	&	$(1.238 \pm 0.001) \times 10^{-3}$	&	$0.71 \pm 0.02$	\\
$1$ &  $\sigma = 0.06$ 	&	$\varepsilon_c = 279.642 \pm 0.001$	&	$(1.484 \pm 0.001) \times 10^{-3}$	&	$0.70 \pm 0.01$	\\
$1$ &  $\sigma = 0.07$ 	&	$\varepsilon_c = 239.692 \pm 0.001$	&	$(1.732 \pm 0.001) \times 10^{-3}$	&	$0.70 \pm 0.02$	\\
$1$ &  $\varepsilon = 250$ 	&	$\sigma_c = 0.0671138 \pm 0.0000002$	&	$(1.660 \pm 0.002) \times 10^{-3}$	&	$0.68 \pm 0.04$	\\
\hline
$2$ &  $\sigma = 0.05$ 	&	$\varepsilon_c = 251.10 \pm 0.01$	&	$(7.6 \pm 0.1) \times 10^{-4}$	&	$0.65 \pm 0.07$	\\
$2$ &  $\sigma = 0.06$ 	&	$\varepsilon_c = 209.25 \pm 0.01$	&	$(9.0 \pm 0.1) \times 10^{-4}$	&	$0.64 \pm 0.10$	\\
$2$ &  $\sigma = 0.07$ 	&	$\varepsilon_c = 179.36 \pm 0.01$	&	$(1.05 \pm 0.02) \times 10^{-3}$	&	$0.67 \pm 0.07$	\\
\hline\hline
\end{tabular}
\caption{Values from the fittings of Fig.~\ref{massgap-fits} to the power law: $M_{\mathrm{AH}}-M_{g} \propto (p_c - p)^\xi$. \label{data_crit}}
\end{center}
\end{table*}

\begin{figure}[b]
\centering
\includegraphics[width=0.4\textwidth]{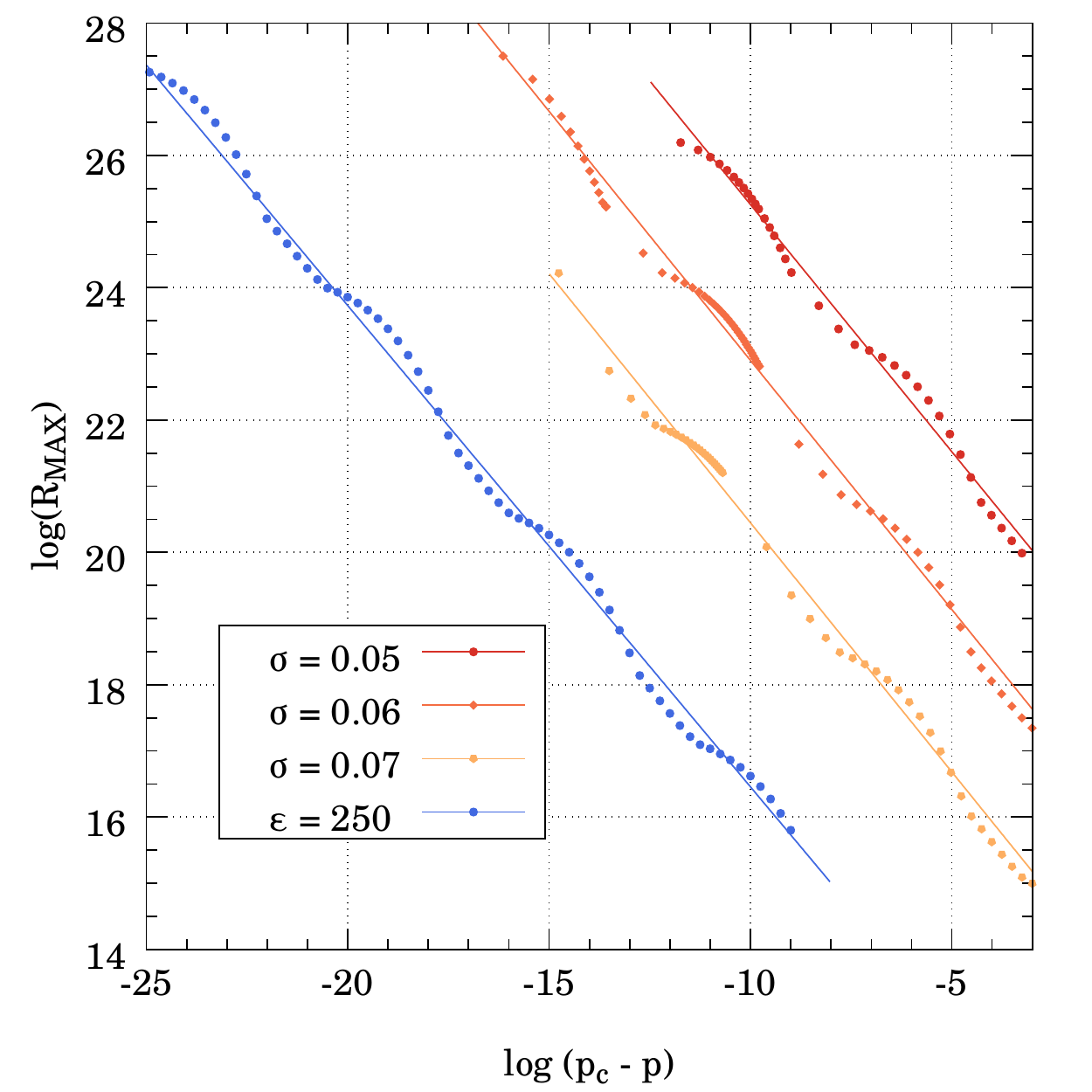}
\caption{Scaling for the curvature scalar for subcritical configurations near the critical point ($p \lesssim p_c$). }
\label{plot_fit_g0}
\end{figure}

We have computed the exponent of the power law, $\xi$, in several ways.  First of all, in a series of simulations for the one- and two-bounce branches, where the initial scalar field width is fixed to the values $\sigma = 0.05, \, 0.06$, and $0.07$.  The results are shown in the left and right plots of Fig.~\ref{massgap-fits}.  Second, in a series of simulations for the one bounce branch at fixed initial scalar field amplitude $\varepsilon = 250$.  The results are shown in the central plot of Fig.~\ref{massgap-fits}.  In Table~\ref{data_crit} we show the results for the different fittings from where it is clear the power law scaling.
We find that all the cases are compatible with a value of $\xi \simeq 0.70$ (note that the precision in the simulations done for the branch of two bounces is inferior due to the higher numerical precision that it requires).  Apart from the mass gap and the power law, these simulations also allow us to conjecture that the exponent is universal in the sense that it is the same for all branches in the diagram of Fig.~\ref{plot_2d_general} and independent of the family of initial conditions: We have shown that it is the same whenever we vary the amplitude or the width and we expect the same to be true for other families of initial data different from the one in Eq.~(\ref{initial-configurations}).

Although the main focus of this study is the behavior of the AH mass when we approach the critical point from the left we can also use our simulations to obtain information about the critical behavior for the cases with $p > p_c$. As it was shown in Ref.~\cite{Garfinkle:1998va}, for subcritical solutions, the maximum value of the Ricci curvature scalar $R=\met^{\mu\nu}R_{\mu\nu}$ at the origin should follow a power law: $R_{max}(x=0) \propto (p_c - p)^{-2\gamma}$ with wiggles in top of that of period $\Delta/2\gamma$ where $\gamma$ and $\Delta$ are, respectively, the critical exponent and the echoing period corresponding to the previous branch.  This should apply also to our subcritical configurations in the near-collapse stage.  We have made a series of simulations for the first critical point for fixed width ($\sigma = 0.05,\, 0.06$, and $0.07$) and for fixed amplitude ($\varepsilon = 250$).  The results are shown in Fig.~\ref{plot_fit_g0} together with the corresponding fitting lines.  Combining the results of all these cases we obtain a value for the critical exponent of $\gamma = 0.373 \pm 0.006$ and an echoing period of $\Delta = 3.443 \pm 0.004$. Notice that the precision obtained for this exponent is significantly better than the one for $\xi$. This is because for this case we can get closer to the critical point given that for computing the maximum of the scalar of curvature, we only need the first part of the evolution until the scalar field starts its trip to the AdS boundary so we can get some extra points besides the simulations used before. As a check,  $\gamma$ and $\Delta$ for the branch of configurations that collapse directly have also been computed prescribing initial conditions on a null hypersurface $u =$ const in our characteristic scheme, obtaining similar results.

To conclude, we have presented the results of a series of Cauchy-characteristic simulations that show the rich structure of AdS spacetime around the multiple critical points of gravitational collapse corresponding to the different bounces of the scalar field at the AdS boundary.  We find that apart from the expected normal type II behavior and scaling of the AH mass for supercritical configurations, we also have a power law for subcritical configurations ($p \lesssim p_c$) of the form $M_{\mathrm{AH}}-M_{g} \propto (p_{c}-p)^{\xi}$ with a minimum AH mass $M_g$ and 
an exponent $\xi\simeq 0.7$ that we conjecture to be the same for all families of initial configurations and also for all the critical points (branches).

\begin{acknowledgements}
We acknowledge support from contracts AYA-2010-15709 (MICINN) and ESP2013-47637-P (MINECO), and high-performance computing resources provided by CSUC and CESGA (projects ICTS-CESGA-249 and ICTS-CESGA-266).  DSO is thankful for support from FPI PhD contract BES-2012-057909 (MINECO).
\end{acknowledgements}



\end{document}